\documentstyle[12pt]{article}
\newcommand{\eq}{\begin{equation}}
\newcommand{\en}{\end{equation}}
\newcommand{\eqn}{\begin{eqnarray}}
\newcommand{\enn}{\end{eqnarray}}

\begin{document}
\begin{titlepage}
\begin{flushright}
hep-th/9402003\\
THEP-94-1 \\
1 February 1994
\end{flushright}
\begin{center}
\LARGE
$SU(3) \times SU(2) \times U(1)$, Higgs, and Gravity \\
 from \\
$Spin(0,8)$ Clifford Algebra $Cl(0,8)$ \\
\vspace{0.5cm}
\large
 Frank D. (Tony) Smith, Jr. \\
\vspace{6pt}

\small
Department of Physics \\
Georgia Institute of Technology \\
Atlanta, Georgia 30332 \\
\vspace{0.5cm}
{\bf Abstract}
\end{center}

The Lagrangian action for the $D_{4}-D_{5}-E_{6}$ model of
hep-th/9306011 \cite{SM1}
\[ \int_{V_{8}} F_{8} \wedge \star F_{8} +
\overline{S^{+}_{8}} \not \! \partial S^{-}_{8} + GF + GH \]
has 8-dim spacetime $V_{8}$ of the vector representation of $Spin(0,8)$

8-dim fermion fields $S^{+}_{8} = S^{-}_{8}$ of the half-spinor reps
of $Spin(0,8)$ and

28 gauge boson fields $F_{8}$ of the bivector adjoint rep of $Spin(0,8)$ .

\vspace{12pt}

In this paper, the structure of the positive definite Clifford algebra
$Cl(0,8)$
of $Spin(0,8)$, and the triality automorphism $V_{8} = S^{+}_{8} = S^{-
}_{8}$,
are used to reduce the spacetime to 4 dimensions and thereby change the
 gauge group from $Spin(0,8)$ to the realistic $SU(3) \times SU(2) \times
U(1)$, Higgs, and Gravity.

The effect of dimensional reduction on fermions, \\
to introduce 3 generations, has been described in hep-ph/9301210 \cite{SM2}.

The global geometry of manifolds $V_{8} = S^{+}_{8} = S^{-}_{8}  =
 {\bf{R}}P^{1} \times S^{7}$,
the effects of dimensional reduction on them, and the calculation of
force strength constants, has been described in hep-th/9302030 \cite{SM3}.

\vspace{12pt}
\normalsize
\footnoterule
\noindent
{\footnotesize \copyright 1994 Frank D. (Tony) Smith, Jr.,
341 Blanton Road, Atlanta, Georgia 30342 USA \\
P. O. Box for snail-mail:  P. O. Box 430, Cartersville,
Georgia 30120 USA \\
e-mail: gt0109e@prism.gatech.edu
and fsmith@pinet.aip.org}
\end{titlepage}

\newpage

\setcounter{footnote}{0}
\section{Introduction}
\setcounter{equation}{0}

The Lagrangian action for the $D_{4}-D_{5}-E_{6}$ model of
hep-th/9306011 \cite{SM1}
\[ \int_{V_{8}} F_{8} \wedge \star F_{8} +
\overline{S^{+}_{8}} \not \! \partial S^{-}_{8} + GF + GH \]
has 8-dim spacetime $V_{8}$ of the vector representation of $Spin(0,8)$

8-dim fermion fields $S^{+}_{8} = S^{-}_{8}$ of the half-spinor reps of
$Spin(0,8)$ and

28 gauge boson fields $F_{8}$ of the bivector adjoint rep of $Spin(0,8)$ .

\vspace{12pt}

In this paper, the structure of the positive definite Clifford algebra
$Cl(0,8)$ of $Spin(0,8)$, and the triality automorphism
$V_{8} = S^{+}_{8} = S^{-}_{8}$, are used to reduce the spacetime to
4 dimensions and thereby change the gauge group from $Spin(0,8)$ to
the realistic $SU(3) \times SU(2) \times U(1)$, Higgs, and Gravity.

\vspace{12pt}

This paper is an extension of a series of papers \cite{SM1, SM2, SM3, SM4}
attempting to construct a realistic model of particle physics and gravity
using structures related to Spin(0,8), the unique Lie algebra with a
triality automorphism between its vector representation and each of its two
half-spinor representations.  All three of those representations are
8-dimensional, and octonionic structures seem to be at the heart of the
special structures used in building the model.

Similar octonionic structures have appeared in other areas of mathematics
and physics, and some of the people working with them attended a meeting
in January 1994 at the Institute for Theoretical Physics at Chalmers
University, G\"{o}teborg, Sweden, hosted by Martin Cederwall and organized
by Geoffrey Dixon.
Octonionic structures in superstrings were presented by Martin Cederwall
and Corinne Manogue; Division algebra structures were presented by
Geoffrey Dixon; and Clifford algebra structures were presented by
Rafal Ablamowicz,  Pertti Lounesto, and Ian Porteous.  Just prior to the
Chalmers meeting, my thinking was influenced by discussions in L\"{u}beck
with Wolfgang Mantke.

This paper is my attempt to use their ideas to further the construction
of the Spin(0,8) model.  To the extent that this paper is useful, credit
should go to them.  However, they should not be blamed for wrong or
useless material in this paper.

\pagebreak

An outline of the  path from $Spin(0,8)$ to $SU(3) \times SU(2) \times
U(1)$, Higgs, and Gravity is the following table of contents:

\vspace{12pt}

2.  $Spin(0,8)$ and its Clifford Algebra $Cl(0,8)$

2.1  Triality and  Half-spinor - Vector Supersymmetry

2.2  $Cl_{e}(0,8)$ Clifford Even Subalgebra  $Cl(0,6) + Cl(0,6)$

\vspace{12pt}

3.  $Cl(0,6)$ = ${\bf{R}}(8)$

\vspace{12pt}

4.  $Cl_{e}(0,6)$ and Dimensional Reduction

4.1  ${\bf{Re}}({\bf{C}}(4))$ - Gravity and Dirac  Complexification

4.2  ${\bf{Im}}({\bf{C}}(4))$ - $SU(3) \times SU(2) \times U(1)$ plus Higgs

\vspace{12pt}

{\bf Some references to material used in this paper are:}

The general structure of Clifford algebras is described in the book of
Ian Porteous \cite{IP}.  A new third edition of his book should
be out later this year.  His book not only describes Clifford algebras,
but is also a good introduction to division algebras and the geometric
actions of $Spin$ groups on spheres.

The papers of Geoffrey Dixon \cite{DIX} are particularly useful
references to division algebras, including the use of division
algebras with respect to spinor spaces of $Spin$ groups.

The global structure of $Spin$ groups is well covered in the paper
of Corinne Manogue and J\"{o}rge Schray \cite{CM}.

The torsion structure of $S^{7}$ and octonions, leading to a
generalized $S^{7}$ algebra, is given by Martin Cederwall and
Christian Preitschopf in \cite{MC1}.

After dimensional reduction to a 4-dimensional spacetime,
structures appear that seem to be related to twistors and
to Hestenes spinors.

A good general reference to twistors is the book of R. O. Wells, Jr.,
\cite{WEL}.  Rafal Ablamowicz relates twistors to Clifford algebras
in \cite{RA}.  Martin Cederwall discusses twistors in \cite{MC2}.

Pertti Lounesto describes Hestenes spinors and Clifford algebras
in \cite{PL}.

\newpage

\section{$Spin(0,8)$ and its Clifford Algebra $Cl(0,8)$}

Begin with the unique triality situation of a $Spin(0,8)$
gauge group, an 8-dimensional spacetime $V_{8}$, and two mirror
image 8-dimensional fermion half-spinor spaces $S^{+}_{8}$ and $S^{-}_{8}$.

They are all contained in the 256-dim positive definite Clifford algebra
$Cl(0,8) = {\bf{R}}(16)$, which has the graded algebra structure:

\[
\begin{array}{ccccccccccccccccc}
1&&8&&28&&56&&70&&56&&28&&8&&1\\
\end{array}
\]

\[
\begin{array}{|cccccccc|cccccccc|}
\hline
0 & 2 & 2 & 2 & 2 & 2 & 2 & 2 & 7 & 5 & 5 & 5 & 5 & 5 & 5 & 5  \\
4 & 4 & 2 & 2 & 2 & 2 & 2 & 2 & 5 & 7 & 5 & 5 & 5 & 5 & 5 & 5  \\
4 & 4 & 4 & 2 & 2 & 2 & 2 & 2 & 5 & 5 & 7 & 5 & 5 & 5 & 5 & 5  \\
4 & 4 & 4 & 4 & 2 & 2 & 2 & 2 & 5 & 5 & 5 & 7 & 5 & 5 & 5 & 5  \\
4 & 4 & 4 & 4 & 4 & 2 & 2 & 2 & 5 & 5 & 5 & 5 & 7 & 5 & 5 & 5  \\
4 & 4 & 4 & 4 & 4 & 4 & 2 & 2 & 5 & 5 & 5 & 5 & 5 & 7 & 5 & 5  \\
4 & 4 & 4 & 4 & 4 & 4 & 4 & 2 & 5 & 5 & 5 & 5 & 5 & 5 & 7 & 5  \\
4 & 4 & 4 & 4 & 4 & 4 & 4 & 4 & 5 & 5 & 5 & 5 & 5 & 5 & 5 & 7  \\
\hline
1 & 3 & 3 & 3 & 3 & 3 & 3 & 3 & 4 & 4 & 4 & 4 & 4 & 4 & 4 & 4  \\
3 & 1 & 3 & 3 & 3 & 3 & 3 & 3 & 6 & 4 & 4 & 4 & 4 & 4 & 4 & 4  \\
3 & 3 & 1 & 3 & 3 & 3 & 3 & 3 & 6 & 6 & 4 & 4 & 4 & 4 & 4 & 4  \\
3 & 3 & 3 & 1 & 3 & 3 & 3 & 3 & 6 & 6 & 6 & 4 & 4 & 4 & 4 & 4  \\
3 & 3 & 3 & 3 & 1 & 3 & 3 & 3 & 6 & 6 & 6 & 6 & 4 & 4 & 4 & 4  \\
3 & 3 & 3 & 3 & 3 & 1 & 3 & 3 & 6 & 6 & 6 & 6 & 6 & 4 & 4 & 4  \\
3 & 3 & 3 & 3 & 3 & 3 & 1 & 3 & 6 & 6 & 6 & 6 & 6 & 6 & 4 & 4  \\
3 & 3 & 3 & 3 & 3 & 3 & 3 & 1 & 6 & 6 & 6 & 6 & 6 & 6 & 6 & 8  \\
\hline
\end{array}
\]

The $GF$ and $GH$ terms of the 8-dimensional Lagrangian
\[ \int_{V_{8}} F_{8} \wedge \star F_{8} +
\overline{S^{+}_{8}} \not \! \partial S^{-}_{8} + GF + GH \]
are the gauge fixing and ghost terms required for quantization.

This paper will deal with the classical terms of the 8-dimensional

\newpage

Lagrangian.  There, the relevant parts of $Cl(0,8)$ are:

\vspace{12pt}

the 1-dim 0-vector scalar part, which will be related to the Higgs scalar;

\vspace{12pt}

the 8-dim 1-vector vector part, which is the 8-dim spacetime $V_{8}$;

\vspace{12pt}

the 28-dim 2-vector bivector part, which gives the 8-dim 28 gauge bosons
of the curvature term $F_{8} \wedge \star F_{8}$, where $\star$ is the
Hodge dual taking the k-vector part into the (8-k)-vector part; and

\vspace{12pt}

the two 8-dim half-spinor parts, which give the 8 first generation
fermion particles and antiparticles.

\vspace{12pt}

The 8-dim half-spinors are defined with respect to the even subalgebra
$Cl_{e}(0,8)$, which contains the even grade parts of $Cl(0,8)$:

\[
\begin{array}{ccccccccccccccccc}
1&&&&28&&&&70&&&&28&&&&1\\
\end{array}
\]

\[
\begin{array}{|cccccccc|cccccccc|}
\hline
0 & 2 & 2 & 2 & 2 & 2 & 2 & 2 &  &  &  &  &  &  &  &   \\
4 & 4 & 2 & 2 & 2 & 2 & 2 & 2 &  &  &  &  &  &  &  &   \\
4 & 4 & 4 & 2 & 2 & 2 & 2 & 2 &  &  &  &  &  &  &  &   \\
4 & 4 & 4 & 4 & 2 & 2 & 2 & 2 &  &  &  &  &  &  &  &   \\
4 & 4 & 4 & 4 & 4 & 2 & 2 & 2 &  &  &  &  &  &  &  &   \\
4 & 4 & 4 & 4 & 4 & 4 & 2 & 2 &  &  &  &  &  &  &  &   \\
4 & 4 & 4 & 4 & 4 & 4 & 4 & 2 &  &  &  &  &  &  &  &   \\
4 & 4 & 4 & 4 & 4 & 4 & 4 & 4 &  &  &  &  &  &  &  &   \\
\hline
 &  &  &  &  &  &  &  & 4 & 4 & 4 & 4 & 4 & 4 & 4 & 4  \\
 &  &  &  &  &  &  &  & 6 & 4 & 4 & 4 & 4 & 4 & 4 & 4  \\
 &  &  &  &  &  &  &  & 6 & 6 & 4 & 4 & 4 & 4 & 4 & 4  \\
 &  &  &  &  &  &  &  & 6 & 6 & 6 & 4 & 4 & 4 & 4 & 4  \\
 &  &  &  &  &  &  &  & 6 & 6 & 6 & 6 & 4 & 4 & 4 & 4  \\
 &  &  &  &  &  &  &  & 6 & 6 & 6 & 6 & 6 & 4 & 4 & 4  \\
 &  &  &  &  &  &  &  & 6 & 6 & 6 & 6 & 6 & 6 & 4 & 4  \\
 &  &  &  &  &  &  &  & 6 & 6 & 6 & 6 & 6 & 6 & 6 & 8  \\
\hline
\end{array}
\]

The two $8 \times 8$ parts of $Cl_{e}(0,8)$ are the +1 and -1
eigenvalue parts of $Cl_{e}(0,8)$ with respect to Clifford multiplication
by the 1-dimensional 8-vector volume pseudoscalar.

The + and - parts each have 8-dimensional column vector
minimal left ideals.  They are isomorphic, so either one (say, +)
can be used as a basis for discussion in this paper.  Each can be
given an octonionic basis $\{ 1, e_{1}, e_{2}, e_{3}, e_{4}, e_{5}, e_{6},
e_{7} \}$ .
They represent the 8-dimensional + and - half-spinor representations
that give the first generation fermion particles and anti-particles.

\vspace{12pt}

The + and - parts each also have 8-dimensional row vector
minimal right ideals.  Each can be given an octonionic basis
$\{ 1, i, j, k, e, ie, je, ke \}$ .  They represent the 8-dimensional
gammas used, along with the covariant derivative, to define
the Dirac operator $\not \! \partial$ .

\[
\begin{array}{|c||cccccccc|}
\hline
 & 1 & i & j & k & e & ie & je & ke  \\
\hline
\hline
1 & 0 & 2 & 2 & 2 & 2 & 2 & 2 & 2  \\
e_{1} & 4 & 4 & 2 & 2 & 2 & 2 & 2 & 2  \\
e_{2} & 4 & 4 & 4 & 2 & 2 & 2 & 2 & 2  \\
e_{3} & 4 & 4 & 4 & 4 & 2 & 2 & 2 & 2  \\
e_{4} & 4 & 4 & 4 & 4 & 4 & 2 & 2 & 2  \\
e_{5} & 4 & 4 & 4 & 4 & 4 & 4 & 2 & 2  \\
e_{6} & 4 & 4 & 4 & 4 & 4 & 4 & 4 & 2  \\
e_{7} & 4 & 4 & 4 & 4 & 4 & 4 & 4 & 4  \\
\hline
\end{array}
\]

\newpage

\subsection{Triality and Half-spinor - Vector Supersymmetry}

Of the physically relevant parts of $Cl(0,8) = {\bf{R}}(16)$,
only the 8-dimensional 1-vector vector spacetime $V_{8}$ part
is not in the even subalgebra  $Cl_{e}(0,8)$.  However, by the
triality automorphism among $V_{8}$, $S^{+}_{8}$ and $S^{-}_{8}$,
all the physically relevant parts of $Cl(0,8) = {\bf{R}}(16)$ can
be described by the even subalgebra  $Cl_{e}(0,8)$.

\[
\begin{array}{|cccccccc|cccccccc|}
\hline
0 & 2 & 2 & 2 & 2 & 2 & 2 & 2 &  &  &  &  &  &  &  &   \\
4 & 4 & 2 & 2 & 2 & 2 & 2 & 2 &  &  &  &  &  &  &  &   \\
4 & 4 & 4 & 2 & 2 & 2 & 2 & 2 &  &  &  &  &  &  &  &   \\
4 & 4 & 4 & 4 & 2 & 2 & 2 & 2 &  &  &  &  &  &  &  &   \\
4 & 4 & 4 & 4 & 4 & 2 & 2 & 2 &  &  &  &  &  &  &  &   \\
4 & 4 & 4 & 4 & 4 & 4 & 2 & 2 &  &  &  &  &  &  &  &   \\
4 & 4 & 4 & 4 & 4 & 4 & 4 & 2 &  &  &  &  &  &  &  &   \\
4 & 4 & 4 & 4 & 4 & 4 & 4 & 4 &  &  &  &  &  &  &  &   \\
\hline
1 &  &  &  &  &  &  &  & 4 & 4 & 4 & 4 & 4 & 4 & 4 & 4  \\
 & 1 &  &  &  &  &  &  & 6 & 4 & 4 & 4 & 4 & 4 & 4 & 4  \\
 &  & 1 &  &  &  &  &  & 6 & 6 & 4 & 4 & 4 & 4 & 4 & 4  \\
 &  &  & 1 &  &  &  &  & 6 & 6 & 6 & 4 & 4 & 4 & 4 & 4  \\
 &  &  &  & 1 &  &  &  & 6 & 6 & 6 & 6 & 4 & 4 & 4 & 4  \\
 &  &  &  &  & 1 &  &  & 6 & 6 & 6 & 6 & 6 & 4 & 4 & 4  \\
 &  &  &  &  &  & 1 &  & 6 & 6 & 6 & 6 & 6 & 6 & 4 & 4  \\
 &  &  &  &  &  &  & 1 & 6 & 6 & 6 & 6 & 6 & 6 & 6 & 8  \\
\hline
\end{array}
\]

As is apparent from the diagram, the spacetime 1-vector $V_{8}$ can
be described through triality by the same row-vector basis
$\{ 1, i, j, k, e, ie, je, ke \}$ used to represent the 8-dimensional
gammas in defining the Dirac operator $\not \! \partial$ .

\vspace{12pt}

{\bf The triality representation of the vector spacetime is effectively
a generalized supersymmetry between vector spacetime and the
half-spinor fermions.  As discussed in \cite{SM1}, it can be extended
by the relationship between vectors and bivectors to a generalized
supersymmetry between fermions and gauge bosons.  It is the reason

\newpage

that $Spin(0,8)$ structures are, in my opinion, the most useful
structures in constructing physically realistic particle physics models.}

\[
\begin{array}{ccccccccccccccccc}
1&&&&28&&&&35&&&&&&&&\\
\end{array}
\]

\[
\begin{array}{|c||cccccccc|}
\hline
 & 1 & i & j & k & e & ie & je & ke  \\
\hline
\hline
1 & 0 & 2 & 2 & 2 & 2 & 2 & 2 & 2  \\
e_{1} & 4 & 4 & 2 & 2 & 2 & 2 & 2 & 2  \\
e_{2} & 4 & 4 & 4 & 2 & 2 & 2 & 2 & 2  \\
e_{3} & 4 & 4 & 4 & 4 & 2 & 2 & 2 & 2  \\
e_{4} & 4 & 4 & 4 & 4 & 4 & 2 & 2 & 2  \\
e_{5} & 4 & 4 & 4 & 4 & 4 & 4 & 2 & 2  \\
e_{6} & 4 & 4 & 4 & 4 & 4 & 4 & 4 & 2  \\
e_{7} & 4 & 4 & 4 & 4 & 4 & 4 & 4 & 4  \\
\hline
\end{array}
\]

\newpage

\subsection{$Cl_{e}(0,8)$ Clifford Even Subalgebra}

The 128-dim even Clifford subalgebra
$Cle(0,8) = {\bf{R}}(16) + {\bf{R}}(16) = Cl(0,7)$
is made up of the sum of two 64-dimensional  Clifford algebras:

${\bf{R}}(16) + {\bf{R}}(16) = Cl(0,6) + Cl(0,6)$ .

One of the two 64-dimensional ${\bf{R}}(16) = Cl(0,6)$ Clifford algebras
is the + half-spinor representation of $Spin(0,8)$, and the other
is the - half-spinor representation.

By the triality automorphism, it is sufficient to discuss either
one of them, say, the + half-spinor of $Spin(0,8)$, which will be
the 64-dimensional

${\bf{R}}(16) = Cl(0,6)$ Clifford algebra discussed here.

As the + part of  $Cl_{e}(0,8)$, it has the graded structure:

\[
\begin{array}{ccccccccccccccccc}
1&&&&28&&&&35&&&&&&&&\\
\end{array}
\]

\[
\begin{array}{|c||cccccccc|}
\hline
 & 1 & i & j & k & e & ie & je & ke  \\
\hline
\hline
1 & 0 & 2 & 2 & 2 & 2 & 2 & 2 & 2  \\
e_{1} & 4 & 4 & 2 & 2 & 2 & 2 & 2 & 2  \\
e_{2} & 4 & 4 & 4 & 2 & 2 & 2 & 2 & 2  \\
e_{3} & 4 & 4 & 4 & 4 & 2 & 2 & 2 & 2  \\
e_{4} & 4 & 4 & 4 & 4 & 4 & 2 & 2 & 2  \\
e_{5} & 4 & 4 & 4 & 4 & 4 & 4 & 2 & 2  \\
e_{6} & 4 & 4 & 4 & 4 & 4 & 4 & 4 & 2  \\
e_{7} & 4 & 4 & 4 & 4 & 4 & 4 & 4 & 4  \\
\hline
\end{array}
\]

It is important to see that, of the + part of  $Cl_{e}(0,8)$, only

the 1-dim 0-vector part, the precursor of the Higgs scalar;

the 28-dim 2-vector part, the $Spin(0,8)$ gauge bosons,

the 8-dim half-spinor column vectors with octonionic basis

$\{ 1, e_{1}, e_{2}, e_{3}, e_{4}, e_{5}, e_{6}, e_{7} \}$; and

the 8-dim spacetime (by triality) row vectors with

octonionic basis $\{ 1, i, j, k, e, ie, je, ke \}$

have physical significance in the Lagrangian.

\vspace{12pt}

The 35-dimensional 4-vector part is redundant.

It goes away naturally by considering $Cl(0,6)$ and its even Clifford
subalgebra.

\vspace{12pt}

First, consider that 28-dimensional  $Spin(0,8)$ is
naturally represented by the upper triangular set of $2$ elements
on the diagram of the + part of $Cl_{e}(0,8)$.

Then, consider that $Spin(0,8)$ has a natural 16-dimensional $U(4)$
subgroup.  If $U_{4}$ denotes an element of $U(4)$, it can be represented
(see section 412 G of \cite{EDM}) as an element of $Spin(0,8)$ by

\[
\begin{array}{|c|c|}
\hline
{\bf{Re}}(U_{4}) &{\bf{Im}}(U_{4})  \\
\hline
{\bf{-Im}}(U_{4}) & {\bf{Re}}(U_{4})  \\
\hline
\end{array}
\]

Then, the $U(4)$ subgroup of $Spin(0,8)$ can be represented
on the diagram of the + part of $Cl_{e}(0,8)$ as

\[
\begin{array}{|c||cccccccc|}
\hline
 & 1 & i & j & k & e & ie & je & ke  \\
\hline
\hline
1 & 0 & U_{4} & U_{4} & U_{4} & U_{4} & U_{4} & U_{4} & U_{4}  \\
e_{1} & 4 & 4 & U_{4} & U_{4} & 2 & U_{4} & U_{4} & U_{4}  \\
e_{2} & 4 & 4 & 4 & U_{4} & 2 & 2 & U_{4} & U_{4}  \\
e_{3} & 4 & 4 & 4 & 4 & 2 & 2 & 2 & U_{4}  \\
e_{4} & 4 & 4 & 4 & 4 & 4 & 2 & 2 & 2  \\
e_{5} & 4 & 4 & 4 & 4 & 4 & 4 & 2 & 2  \\
e_{6} & 4 & 4 & 4 & 4 & 4 & 4 & 4 & 2  \\
e_{7} & 4 & 4 & 4 & 4 & 4 & 4 & 4 & 4  \\
\hline
\end{array}
\]

Then, by moving some $U_{4}$ elements on or below the
diagonal of the diagram of the + part of $Cl_{e}(0,8)$,
and moving the 0-vector and some 4-vector elements above
the diagonal:

\[
\begin{array}{|c||cccccccc|}
\hline
 & 1 & i & j & k & e & ie & je & ke  \\
\hline
\hline
1 & U_{4} & U_{4} & U_{4} & U_{4} & 0 & 4 & 4 & 4  \\
e_{1} & U_{4} & U_{4} & U_{4} & U_{4} & 2 & 4 & 4 & 4  \\
e_{2} & U_{4} & U_{4} & U_{4} & U_{4} & 2 & 2 & 4 & 4  \\
e_{3} & U_{4} & U_{4} & U_{4} & U_{4} & 2 & 2 & 2 & 4  \\
e_{4} & 4 & 4 & 4 & 4 & 4 & 2 & 2 & 2  \\
e_{5} & 4 & 4 & 4 & 4 & 4 & 4 & 2 & 2  \\
e_{6} & 4 & 4 & 4 & 4 & 4 & 4 & 4 & 2  \\
e_{7} & 4 & 4 & 4 & 4 & 4 & 4 & 4 & 4  \\
\hline
\end{array}
\]

Then, by moving the 0-vector and the upper 2-vector elements
on or below the diagonal of the diagram of the + part of $Cl_{e}(0,8)$,
and moving some 4-vector elements above the diagonal:

\[
\begin{array}{|c||cccccccc|}
\hline
 & 1 & i & j & k & e & ie & je & ke  \\
\hline
\hline
1 & U_{4} & U_{4} & U_{4} & U_{4} & 4 & 4 & 4 & 4  \\
e_{1} & U_{4} & U_{4} & U_{4} & U_{4} & 4 & 4 & 4 & 4  \\
e_{2} & U_{4} & U_{4} & U_{4} & U_{4} & 4 & 4 & 4 & 4  \\
e_{3} & U_{4} & U_{4} & U_{4} & U_{4} & 4 & 4 & 4 & 4  \\
e_{4} & 4 & 4 & 4 & 4 & 2 & 2 & 2 & 2  \\
e_{5} & 4 & 4 & 4 & 4 & 2 & 2 & 2 & 2  \\
e_{6} & 4 & 4 & 4 & 4 & 2 & 2 & 2 & 2  \\
e_{7} & 4 & 4 & 4 & 4 & 4 & 4 & 4 & 0  \\
\hline
\end{array}
\]

Then, since the 4-vector elements are not useful in the physical model,
eliminate them to get the physical part of the diagram of
the 64-dimensional + part of $Cl_{e}(0,8)$:

\[
\begin{array}{|c||cccccccc|}
\hline
 & 1 & i & j & k & e & ie & je & ke  \\
\hline
\hline
1 & U_{4} & U_{4} & U_{4} & U_{4} &  &  &  &     \\
e_{1} & U_{4} & U_{4} & U_{4} & U_{4} &  &  &  &     \\
e_{2} & U_{4} & U_{4} & U_{4} & U_{4} &  &  &  &     \\
e_{3} & U_{4} & U_{4} & U_{4} & U_{4} &  &  &  &     \\
e_{4} &  &  &  &  & 2 & 2 & 2 & 2    \\
e_{5} & &  &  &  & 2 & 2 & 2 & 2    \\
e_{6} &  &  &  &  & 2 & 2 & 2 & 2    \\
e_{7} &  &  &  &  &  &  &  & 0    \\
\hline
\end{array}
\]

The physical part of the + part of $Cl_{e}(0,8)$ is a 29-dimensional
subspace of the 32-dimensional block diagonal subspace of
the 64-dimensional + part of $Cl_{e}(0,8)$ with two
16-dimensional square blocks.

The physical part is one 16-dimensional block that is $U(4) \subset Spin(0,8)$,
and a 13-dimensional subspace of the other 16-dimensional block that
is the 0-vector $Spin(0,8)$ scalar plus the remaining 12 = 28 - 16
dimensions of the 2-vector adjoint bivectors of $Spin(0,8)$.

\vspace{12pt}

Since $U(4) / U(1) = SU(4)$, $SU_{4}$ = $Spin(0,6)$, and
the + part of $Cl_{e}(0,8)$ = $Cl(0,6)$, it is natural at this point
to look at $Cl(0,6)$.

\newpage

\section{$Cl(0,6)$ = ${\bf{R}}(8)$}

The + part of $Cl_{e}(0,8)$ = $Cl(0,6)$, the positive definite Clifford
algebra of $Spin(0,6)$ = $SU(4)$ $\subset U(4)$.

$Cl(0,6)$ is 64-dimensional, and has the graded structure

\[
\begin{array}{ccccccccccccccccc}
&&1&&6&&15&&20&&15&&6&&1&&\\
\end{array}
\]

\[
\begin{array}{|c||cccc|cccc|}
\hline
 & 1 & i & j & k & e & ie & je & ke  \\
\hline
\hline
1 & 0 & 2 & 2 & 2 & 3 & 5 & 5 & 5 \\
e_{1} & 2 & 2 & 2 & 2 & 3 & 3 & 5 & 5 \\
e_{2} & 2 & 2 & 2 & 2 & 3 & 3 & 3 & 5 \\
e_{3} & 2 & 2 & 2 & 2 & 3 & 3 & 3 & 3 \\
\hline
e_{4} & 3 & 3 & 3 & 3 & 4 & 4 & 4 & 4 \\
e_{5} & 1 & 3 & 3 & 3 & 4 & 4 & 4 & 4 \\
e_{6} & 1 & 1 & 3 & 3 & 4 & 4 & 4 & 4 \\
e_{7} & 1 & 1 & 1 & 3 & 4 & 4 & 4 & 6 \\
\hline
\end{array}
\]

Just as with the + part of $Cl_{e}(0,8)$, $Cl(0,6)$ can be
given a column vector fermion particle octonionic basis
$\{ 1, e_{1}, e_{2}, e_{3}, e_{4}, e_{5}, e_{6}, e_{7} \}$
and a row vector spacetime octonionic basis
$\{ 1, i, j, k, e, ie, je, ke \}$.

\vspace{12pt}

The 2-vector bivector part of $Cl(0,6)$ is just $SU(4) \subset U(4)$,
and, if the 0-vector part of $Cl(0,6)$ is taken to be $U(1) \subset U(4)$,
the upper 16-dimensional block diagonal is the same $U(4)$ that is
a subgroup of $Spin(0,8)$ in the + part of $Cl_{e}(0,8)$.

\vspace{12pt}

The lower 16-dimensional block diagonal of 15 4-vectors and a 6-vector
can be transformed by the Hodge $\star$ to 15 2-vectors and a 0-vector.

The transformed 0-vector corresponds to the 0-vector scalar of
the + part of $Cl_{e}(0,8)$.

Of the 15 transformed $Cl(0,6)$ 4-vectors, 12 correspond to the
12 remaining $Cl(0,8)$ 2-vectors that are outside the $U(4) \subset
Spin(0,8)$,
and the other 3 are not physically relevant, as they correspond to
$Cl(0,8)$ 4-vectors.

 \vspace{12pt}

As all the physically relevant parts of the + part of $Cl_{e}(0,8)$ are
in the two 16-dimensional block diagonal parts of $Cl(0,6)$, it is now
natural to look at the 32-dimensional even subalgebra $Cl_{e}(0,6)$ of
$Cl(0,6)$.

\newpage

\section{ $Cl_{e}(0,6)$ and Dimensional Reduction}

$Cl_{e}(0,6) = Cl(0,5) =  {\bf{C}}(4) = {\bf{Re}}({\bf{C}}(4)) +
{\bf{Im}}({\bf{C}}(4))$
is 32-dimensional, and has the graded structure:

\[
\begin{array}{ccccccccccccccccc}
&&1&&&&15&&&&15&&&&1&&\\
\end{array}
\]

\[
\begin{array}{|c||cccc|cccc|}
\hline
 & 1 & i & j & k & e & ie & je & ke  \\
\hline
\hline
1 & 0 & 2 & 2 & 2 & & & & \\
e_{1} & 2 & 2 & 2 & 2 & & & & \\
e_{2} & 2 & 2 & 2 & 2 & & & & \\
e_{2} & 2 & 2 & 2 & 2 & & & & \\
\hline
e_{4} &  & & & & 4 & 4 & 4 & 4 \\
e_{5} &  & & & & 4 & 4 & 4 & 4 \\
e_{6} &  & & & & 4 & 4 & 4 & 4 \\
e_{7} &  & & & & 4 & 4 & 4 & 6 \\
\hline
\end{array}
\]

$Cl_{e}(0,6)$ can be given a column vector fermion particle octonionic basis
$\{ 1, e_{1}, e_{2}, e_{3}, e_{4}, e_{5}, e_{6}, e_{7} \}$
and a row vector spacetime octonionic basis
$\{ 1, i, j, k, e, ie, je, ke \}$.

\vspace{12pt}

The 2-vector bivector part of $Cl_{e}(0,6)$ is just $SU(4) \subset U(4)$,
and, if the 0-vector part of $Cl_{e}(0,6)$ is taken to be $U(1) \subset
U(4)$, the upper 16-dimensional block diagonal is the same $U(4)$ that is
a subgroup of $Spin(0,8)$ in the + part of $Cl_{e}(0,8)$.

\vspace{12pt}

The lower 16-dimensional block diagonal of 15 4-vectors and a 6-vector
can be transformed by the Hodge $\star$ to 15 2-vectors and a 0-vector.

The transformed 0-vector corresponds to the 0-vector scalar of
the + part of $Cl_{e}(0,8)$.

Of the 15 transformed $Cl_{e}(0,6)$ 4-vectors, 12 correspond to the
12 remaining $Cl(0,8)$ 2-vectors that are outside the $U(4) \subset
Spin(0,8)$,
and the other 3 are not physically relevant, as they correspond to
$Cl(0,8)$ 4-vectors.

\vspace{12pt}

If $U_{4}$ denotes an element of $U(4) \subset Spin(0,8)$, and if
0 and 2 denote the $Spin(0,8)$ scalar 0-vector and bivector 2-vectors,
a 29-dimensional subspace of 32-dimensional $Cl{e}(0,6)$ can now
be seen to correspond to the physically relevantpart of the  + part
of $Cl_{e}(0,8)$, with graded structure:

\[
\begin{array}{ccccccccccccccccc}
&&1&&&&15&&&&15&&&&1&&\\
\end{array}
\]

\[
\begin{array}{|c||cccccccc|}
\hline
 & 1 & i & j & k & e & ie & je & ke  \\
\hline
\hline
1 & U_{4} & U_{4} & U_{4} & U_{4} &  &  &  &     \\
e_{1} & U_{4} & U_{4} & U_{4} & U_{4} &  &  &  &     \\
e_{2} & U_{4} & U_{4} & U_{4} & U_{4} &  &  &  &     \\
e_{3} & U_{4} & U_{4} & U_{4} & U_{4} &  &  &  &     \\
e_{4} &  &  &  &  & 2 & 2 & 2 & 2    \\
e_{5} & &  &  &  & 2 & 2 & 2 & 2    \\
e_{6} &  &  &  &  & 2 & 2 & 2 & 2    \\
e_{7} &  &  &  &  &  &  &  & 0    \\
\hline
\end{array}
\]

This $8 \times 8$ diagram of $Cl_{e}(0,6)$ is clearly equivalent to the
following
$8 \times 4$ diagram of $Cl(0,5)$ = $Cl_{e}(0,6)$ = ${\bf{C}}(4)$.

\[
\begin{array}{ccccccccccccccccc}
&&&1&&5&&10&&10&&5&&1&&&\\
\end{array}
\]

\[
\begin{array}{|c||cccc|}
\hline
 & 1 & i & j & k  \\
\hline
\hline
1 & U_{4} & U_{4} & U_{4} & U_{4} \\
e_{1} & U_{4} & U_{4} & U_{4} & U_{4} \\
e_{2} & U_{4} & U_{4} & U_{4} & U_{4} \\
e_{3} & U_{4} & U_{4} & U_{4} & U_{4} \\
\hline
\hline
e_{4} & 2 & 2 & 2 & 2 \\
e_{5} & 2 & 2 & 2 & 2 \\
e_{6} & 2 & 2 & 2 & 2 \\
e_{7} &  &  &  & 0 \\
\hline
\end{array}
\]

{\bf The dimension of row vector spacetime has been reduced

 from 8, with basis $\{ 1, i, j, k, e, ie, je, ke \}$,

to 4, with basis $\{ 1, i, j, k \}$.}

This dimensional reduction mechanism, which can be seen
as taking the octonionic basis element $e$ of
$\{ 1, i, j, k, e, ie, je, ke \}$
into $1$, or as ignoring $e$ as unobservable or unphysical,
leaving a quaternionic 4-dimensional spacetime with
basis $\{ 1, i, j, k \}$, has been advocated by Martin Cederwall
and Corinne Manogue with respect to dimensional reduction of
superstring theory from 10 dimensions to 4 dimensions, and
by me \cite{SM2,SM3,SM4} in dimensional reduction of my model
 from 8 dimensions to 4 dimensions.

\vspace{12pt}

After dimensional reduction, there are two 16-dimensional spaces in

$Cl_{e}(0,6) = Cl(0,5) =  {\bf{C}}(4) = {\bf{Re}}({\bf{C}}(4)) +
{\bf{Im}}({\bf{C}}(4))$:

${\bf{Re}}({\bf{C}}(4))$, with 16-dimensional $U(4)$; and

${\bf{Im}}({\bf{C}}(4))$, of which 16-dimensional space only
13 dimensions are physically relevant, the 1-dim $Spin(0,8)$ 0-vector scalar
and the 12-dim remainder of $Spin(0,8)$ outside the subgroup $U(4)$.

\vspace{12pt}

Each of the 16-dimensional spinor spaces may have some relationship
to the operator spinors of Hestenes and Keller that are described by
Pertti Lounesto in \cite{PL}.

\newpage

\subsection{${\bf{Re}}({\bf{C}}(4))$ - Gravity and Dirac  Complexification}

Consider first the ${\bf{Re}}({\bf{C}}(4))$ part of $Cl_{e}(0,6) =
Cl(0,5)$, with 16-dimensional $U(4)$.

As it is in the upper left diagonal block of $Cl(0,6)$, the $U(4)$ gauge
group acts directly on the surviving 4-dimensional spacetime with basis
$\{ 1, i, j, k \}$.  Action on the first-generation fermion particle basis
$\{ 1, e_{1}, e_{2}, e_{3}, e_{4}, e_{5}, e_{6}, e_{7} \}$ is
direct with respect to $\{ 1, e_{1}, e_{2}, e_{3} \}$, and defined
on $\{ e_{4}, e_{5}, e_{6}, e_{7} \}$ by the octonion automorphism
between the two quaternionic subspaces of the octonions.

It should be noted that the $U(4)$ structure, and its related conformal
structure, may be related to twistors.

The resulting structure is:

\[
\begin{array}{ccccccccccccccccc}
&&&1&&&&10&&&&5&&&&\\
\end{array}
\]

\[
\begin{array}{|c|c||cccc|}
\hline
 &  & 1 & i & j & k  \\
\hline
\hline
1 & e_{4} & U_{4} & U_{4} & U_{4} & U_{4} \\
e_{1} & e_{5} & U_{4} & U_{4} & U_{4} & U_{4} \\
e_{2} & e_{6} & U_{4} & U_{4} & U_{4} & U_{4} \\
e_{3} & e_{7} & U_{4} & U_{4} & U_{4} & U_{4} \\
\hline
\end{array}
\]

Since $U(4) / U(1) = SU(4) = Spin(0,6)$,
and $Spin(0,6)$ is the compact version of the
conformal group of 4-dimensional spacetime,
the 15-dimensional $SU(4) = Spin(0,6)$ can be
used as a local gauge group symmetry to produce
the Einstein-Hilbert action for gravity, as has
been shown by Mohapatra in section 14.6 of \cite{MOH}.

Mohapatra's conformal group approach is similar to
the approach of MacDowell and Mansouri \cite{McM}
that I have used in \cite{SM3,SM4}.  MacDowell and Mansouri
used an $Sp(2) = Spin(5)$ local gauge symmetry group to get
Einstein-Hilbert gravity.

Mohapatra used the same method, but used gauge fixing
of the 5 conformal degrees of freedom to reduce
conformal $Spin(0,6)$ to 10-dimensional $Spin(0,5)$,
the bivector Lie algeba of $Cl(0,5)$.

\newpage

The relation of the 5 conformal degrees of freedom to the
10 dimensions of $Spin(0,5)$ is shown in the following
diagram of the even part of $Cl(0,5)$:

\[
\begin{array}{ccccccccccccccccc}
&&&1&&&&10&&&&5&&&&\\
\end{array}
\]

\[
\begin{array}{|c|c||cccc|}
\hline
 &  & 1 & i & j & k  \\
\hline
\hline
1 & e_{4} & 0 & 4 & 4 & 2 \\
e_{1} & e_{5} & 4 & 4 & 2 & 2 \\
e_{2} & e_{6} & 4 & 2 & 2 & 2 \\
e_{3} & e_{7} & 2 & 2 & 2 & 2 \\
\hline
\end{array}
\]

The 10 $Cl(0,5)$ 2-vectors are the $Spin(0,5)$ gauge group
that produces Einstein-Hilbert gravity.

The 5 $Cl(0,5)$ 4-vectors are the conformal degrees of
freedom that are gauge-fixed by Mohapatra in his production
of Einstein-Hilbert gravity.

\vspace{12pt}

{\bf If the 5 conformal degrees of freedom are not fixed, but
are considered to be physical and used, the result is the quantum
theory of gravity described by Narlikar and Padmanabhan in
chapters 12 and 13 of \cite{NP}.}

\vspace{12pt}

{\bf The $Cl(0,5)$ 0-vector is the $U(1) = U(4)/SU(4)$.

Physically, the $U(1)$ is the Dirac complexification and
it gives the physical Dirac gammas their complex structure,
so that they are ${\bf{C}}(4)$ instead of $ {\bf{R}}(4)$.}

\vspace{12pt}

There are two ways to take the even subalgebra of $Cl_{e}(0,6)$.

The compact, positive-definite, Euclidean way

is to use $Cl_{e}(0,6) = {\bf{C}}(4) = Cl(0,5)$.

Then $Cl_{e}(0,5) = Cl(0,4) = {\bf{H}}(2) = Cl(1,3)$.

The non-compact, anti-deSitter, Minkowski way

is to use $Cl_{e}(0,6) = {\bf{C}}(4) = Cl(2,3)$.

Then $Cl_{e}(2,3) = Cl(2,2) = {\bf{R}}(4) = Cl(3,1)$.

\vspace{12pt}

${\bf{C}} \otimes  {\bf{H}}(2) =
 {\bf{C}} \otimes {\bf{R}}(4) = {\bf{C}}(4)$,

{\bf so Dirac complexification justifies the physical use of
Wick rotation between Euclidean and Minkowski spacetimes.}

\newpage

\subsection{${\bf{Im}}({\bf{C}}(4))$ - $SU(3) \times SU(2) \times U(1)$
plus Higgs}

Consider second the ${\bf{Im}}({\bf{C}}(4))$ part of $Cl_{e}(0,6) =
Cl(0,5)$, with a scalar 0-vector corresponding to the 0-vector scalar of
the + part of $Cl_{e}(0,8)$ and

with 12 bivectors corresponding to the 12 remaining $Cl(0,8)$ 2-vectors
that are outside the $U(4) \subset Spin(0,8)$.

There are 3 degrees of freedom in 16-dimensional  ${\bf{Im}}({\bf{C}}(4))$
that are not physically relevant, as they correspond to $Cl(0,8)$ 4-vectors.

As these 13 degrees of freedom are in the lower right diagonal block
of $Cl(0,6)$, they acted directly on the 4-dimensional part of
8-dimensional spacetime with basis $\{ e, ie, je, ke \}$ that did not
survive the dimensional reduction process, called the collapsed dimensions.

Since they did not act directly on the surviving 4-dimensional
spacetime with basis $\{ 1, i, j, k \}$ , their action after dimensional
reduction should be the action of a local gauge symmetry rather than
the action of a spacetime symmetry.

Their action on the first-generation fermion particle basis

$\{ 1, e_{1}, e_{2}, e_{3}, e_{4}, e_{5}, e_{6}, e_{7} \}$ is

direct with respect to $\{ e_{4}, e_{5}, e_{6}, e_{7} \}$, and defined
on $\{ 1, e_{1}, e_{2}, e_{3} \}$ by the octonion automorphism
between the two quaternionic subspaces of the octonions.

The resulting structure is:

\[
\begin{array}{|c|c||cccc|}
\hline
 &  & 1 & i & j & k  \\
\hline
\hline
1 & e_{4} & 2 & 2 & 2 & 2 \\
e_{1} & e_{5} & 2 & 2 & 2 & 2 \\
e_{2} & e_{6} & 2 & 2 & 2 & 2 \\
e_{3} & e_{7} &  &  &  & 0 \\
\hline
\end{array}
\]

The $3 \times 3$ upper left block of bivectors forms a
local $U(3)$ gauge group.

The 3 far right-column bivectors form a local $SU(2)$ gauge group.

The scalar 0-vector forms the Higgs scalar.

\newpage

If the elements of $U(3)$, $SU(2)$, and the Higgs are denoted respectively
by $U_{3}$, $SU_{2}$, and $H$, the following structure results:

\[
\begin{array}{|c|c||cccc|}
\hline
 &  & 1 & i & j & k  \\
\hline
\hline
1 & e_{4} & U_{3} & U_{3} & U_{3} & SU_{2} \\
e_{1} & e_{5} & U_{3} & U_{3} & U_{3} & SU_{2} \\
e_{2} & e_{6} & U_{3} & U_{3} & U_{3} & SU_{2} \\
e_{3} & e_{7} &  &  &  & H \\
\hline
\end{array}
\]

Since $U(1) = U(3) / SU(3)$, the structure is a version of the standard
model

$SU(3) \times SU(2) \times U(1)$ plus Higgs.

Details of this model, including the geometry of the Higgs mechanism and
calculations of force strengths and particle masses, are given in
\cite{SM1,SM2,SM3,SM4}.

\vspace{12pt}

{\bf  The $SU(3) \times SU(2) \times U(1)$ plus Higgs as described in
this section, together with the conformal gravity and Dirac complexification
described in the preceding section, produce a realistic and consistent
model of particle physics plus gravity.}

\end{document}